\documentclass[prd,aps,nofootinbib,preprintnumbers,
showpacs,showkeys]{revtex4}
\usepackage{graphicx,epsf,amsmath,amsfonts,amssymb,amsbsy}
\newcommand{\ds}{\displaystyle}
\newcommand{\vev}[1]{\langle#1\rangle}
\newcommand{\mat}{\left ( \begin{array}}
\newcommand{\emat}{\end{array} \right )}
\newcommand{\vect}{\left ( \begin{array}{c}}
\newcommand{\evect}{\end{array} \right )}

\newcommand{\Proj}[1]{\mathrm{P}\kern-6.5pt\mathrm{P}_{\mathrm{#1}}}
\newcommand{\MatrixUnit}{{1}\kern-4.5pt{1}}

\newcommand{\Ltildeplus}{\widetilde{\Lambda}_+}

\newcommand{\Ltildeminus}{\widetilde{\Lambda}_-}

\newcommand{\Tr}{\mathrm{Tr}}

\newcommand{\Edeltabr}[1]{(E_{\Delta}^{#1})}

\newcommand{\expv}[1]{\langle {#1}\rangle}

\preprint{HU-EP-04/70}

\begin{document}
\title{Pion, $\sigma$-meson and diquarks in the 2SC phase
of dense cold quark matter}
\author{D.~Ebert}
\email{debert@physik.hu-berlin.de}
\affiliation{Institut f\"ur Physik,
Humboldt-Universit\"at zu Berlin, 12489 Berlin, Germany}
\author{K. G.~Klimenko}
\email{kklim@mx.ihep.su}
\affiliation{Institute
of High Energy Physics, 142281, Protvino, Moscow Region, Russia}
\author{V. L. Yudichev}
\email{yudichev@thsun1.jinr.ru}
\affiliation{Joint Institute for Nuclear Research, 141980, Dubna,
Moscow Region, Russia}

\begin{abstract}
The spectrum of  meson and diquark excitations of  dense cold
quark matter is investigated in the framework of a
Nambu--Jona-Lasinio type model for light quarks of two flavors.
It was found out that a first order phase transition occurs when the
chemical potential $\mu$ exceeds the critical value $\mu_c=350$~MeV.
Above $\mu_c$, the diquark condensate $\vev{qq}$ forms, breaking the
color symmetry of strong interaction. The masses of $\pi$- and
$\sigma$-mesons are shown to grow with the chemical potential $\mu$
in the color-superconducting phase, but the mesons themselves become
almost stable particles due to the Mott effect. Moreover, we have
found in this phase an abnormal number of three, instead of five
Nambu--Goldstone bosons, together with a  color doublet of light
stable diquark modes and a color-singlet heavy diquark resonance with
the mass $\sim$~1100 MeV. In the color symmetric phase,
\textit{i.~e.\/} for $\mu <\mu_c$, a mass splitting of diquarks and
antidiquarks is shown to arise if $\mu\ne 0$, contrary to the case of
vanishing chemical potential, where the masses of antidiquarks and
diquarks are degenerate at the value $\sim$~700 MeV.
\end{abstract}

% 11.30.Qc Spontaneous and radiative symmetry breaking
% 12.39.-x Phenomenological quark models
% 11.15.Ex Spontaneous breaking of gauge symmetries
% 12.38.Aw General properties of QCD
% 12.38.Mh Quark-gluon plasma
% 11.10.St Bound and unstable states; Bethe-Salpeter equations
% 12.38.-t Quantum chromodynamics
% 12.38.Lg Other nonperturbative calculations
% 26.60.+c Nuclear matter aspects of neutron stars
% 21.65.+f Nuclear matter
% 12.39.Fe

\maketitle
%\draft
%\large
%\maketitle

\section{Introduction}

One of the challenging problems of elementary particle physics
is the investigation of hot and/or dense strongly interacting matter.
At normal conditions (low temperatures and baryon densities),
it is in the hardronic phase, where quarks and gluons are confined
and chiral symmetry is broken. It is widely believed that at high
temperature strongly interacting matter exists as a quark-gluon
plasma (QGP). Another example of matter under extreme conditions is
the interior of compact stars (low temperature and rather high baryon
densities) which presumably consists of nothing else than
dense cold quark matter. Hopefully, the properties of quark
matter in extreme conditions become observable in relativistic
heavy-ion collision experiments and/or through modifications of star
evolution processes.

Clearly, the underlying theory of strongly
interacting matter, both in the vacuum and in extreme conditions, is
QCD. Unfortunately, a proper and reliable quantitative description of
quark matter in terms of a perturbative expansion of QCD
is available only for asymptotically high values of temperature
and/or chemical potential (densities). Nevertheless, perturbative QCD
calculations show that at high temperature and low baryon density
there is actually a QGP phase, with quarks and gluons being free
particles and with the chiral symmetry being restored. Recently, a
progress has been done in extending the lattice QCD approach to small
nonvanishing values ($\lesssim 140$~MeV) of the chemical potential
$\mu$ (see, \textit{e.~g.\/}\ \cite{Fodor:2004ft} and references
therein).  However, for the range of interest, \textit{i.~e.\/} for
$\mu\sim 300 - 400$~MeV, lattice QCD does not help. Anyway,
perturbative QCD considerations, performed at low temperature and
asymptotically high baryon density (large $\mu$), indicate the
occurrence of a new color superconducting phase of cold quark matter
\cite{son}. The confinement is also absent in it, but the ground
state, unlike the case of QGP, is characterized by a nonvanishing
diquark condensate $\vev{qq}$.

At moderate baryon densities, the QCD coupling constant is too
large so that perturbation theory fails in this case. Obviously, the
investigation of quark matter can then be suitably done in the
framework of an effective quark model, \textit{e.~g.\/}\ in a chiral
quark model of the Nambu--Jona-Lasinio (NJL)  type \cite{njl}
including various channels of four-fermion interactions. Recently, on
the basis of NJL type models, it was shown that a color
superconducting phase might yet be present at rather small values of
$\mu\sim 350$ MeV, \textit{i.~e.\/} for baryon densities just several
times larger than the density of the ordinary nuclear matter (see,
\textit{e.~g.\/}, \cite{alford,neutr} for a review).  But this
density  is presumably reached in the cores of compact stars. On the
other hand, the color superconducting quark matter inside compact
stars, if it exists, can reveal itself through modifications of the
star evolution process. The latter is the subject of astrophysical
observations and might be actually picked out from the data which
are now being collected. By these reasons, the color-superconducting
quark matter surely deserves a more detailed study.

On the microscopic level, the processes running inside compact
stars and in the fireballs created in heavy-ion collisions are
understood as governed, to much extent, by strongly interacting
quarks and gluons. Recent experiments at RHIC have shown that the
hot and dense quark matter reached at the collider is far from full
asymptotic freedom; instead, it is `strongly coupled'
\cite{Shuryak:2004cy}. In this case, the correlations between quarks
and antiquarks, in particular due to composite mesons, are not
negligible. This is why we start an investigation of the
pion and $\sigma$-meson in the quark matter in extreme conditions.
These particles are expected to be numerously produced in
heavy-ion collisions. We also investigate diquarks, because they are
important in determining baryon properties.

In the present paper we work with an extended two-flavor NJL model
in order to study the ground state of cold quark matter and some
lightest meson and diquark excitations in the two-flavor
color superconducting phase (2SC). For simplicity, only a single
quark chemical potential $\mu$ (common for all quarks) is used in the
model. Moreover, we restrict ourselves to the region $\mu\leq
400$~MeV (notice that at higher values of $\mu$ the
color-flavor-locked phase is a more preferable one \cite{alford}).
Insofar as the phase diagram of quark matter has been discussed in
a lot of other papers, we focus on the meson and diquark excitations
only. Partially, we have already addressed this problem in our
previous paper \cite{bekvy}, where we had shown that in the NJL model
under consideration an anomalous number of Nambu--Goldstone bosons
is present when the initial $\rm SU(3)_c$ symmetry of the model is
spontaneously broken down to $\rm SU(2)_c$. Now, we are interested in
the investigation of the masses of the remaining heavy diquark as
well as of the pion and the $\sigma$-meson.

The paper is organized as follows. In Section II, we present the
Lagrangian of the extended NJL  model as well as an equivalent
Lagrangian that contains meson and diquark fields coupled with
quarks. Using the Nambu--Gorkov formalism, we then derive an
expression for the quark propagator, which is suitable for the
study of the 2SC phase. Its poles provide us with the (anti)quark
dispersion relations. Moreover, in this section, the gap equations
for both the chiral and diquark condensates are derived.  The meson
and diquark masses are calculated in Section III; they are found to
be monotonously increasing functions of $\mu$ in the 2SC phase.
Moreover, the pion and $\sigma$-meson become almost stable
particles in the 2SC phase due to the Mott effect
(they are allowed to decay through electro-weak channels only).
Finally, we discuss the mass splitting between the diquark and
antidiquark states arising in the color-symmetric phase and the
phenomenon of an anomalous number of  NG-bosons in the  2SC phase.
Section~IV contains our conclusions and discussion. Some
technical details are worked out in two Appendices.

\section{Two-flavor NJL model}
\subsection{Lagrangian}

In the original version of the NJL model \cite{njl}, the
four-fermion interaction of a proton $(p)$ and neutron $(n)$
doublet was considered, and the principle of the dynamical breaking
of chiral symmetry was demonstrated. Later, the $(p,n)$-doublet
was replaced by a doublet of colored up $(u)$ and down $(d)$ quarks
(or even more generally, by a flavor triplet)
in order to describe phenomenologically the physics of
light mesons \cite{ebvol,volk,hatsuda,volkyud}, diquarks
\cite{ebka,vog} as well as the meson-baryon interaction
\cite{ebjur,reinh}. In this sense, the NJL model may be thought
of as an effective theory for low-energy QCD.
\footnote{Indeed, let us consider two-flavor $\rm SU(3)_c$-symmetric
QCD. By integrating over gluons in the generating functional of QCD
and further ``approximating'' the nonperturbative gluon propagator
by a $\delta$-function, one arrives at an effective local
chiral four-quark interaction of the NJL type, describing
low-energy hadron physics. Moreover, after the Fierz transformation
of the interaction terms, one obtains an NJL-type Lagrangian that
describes the interaction of quarks in the scalar and pseudo-scalar
$(\bar q q)$ as well as in the scalar diquark $(qq)$ channels (see,
\textit{e.~g.\/}, Lagrangian (\ref{1}) in this section).} (Of course,
one should keep in mind that, unlike  QCD, quarks are not confined in
the NJL model.) At present time, the phenomenon of dynamical
(chiral) symmetry breaking is one of the cornerstones of modern
particle physics. It has been studied, for example, in the framework
of NJL-type models with external magnetic fields \cite{mir}, in
curved space-times \cite{odin}, in spaces with nontrivial
topology \cite{incera}, etc. In particular, the properties of
normal hot and/or dense quark matter were also considered within
such models \cite{hatsuda,asakawa,ebert,klim}. NJL-type models
still remain a simple but useful instrument for the exploration
of color superconducting quark matter at moderate densities
\cite{alford,skp,bvy03}, where analytical and/or lattice
computations in  QCD are hindered.

Instead of formulating a thermal theory in Euclidean metric
(which is natural in statistical physics when one wants to derive
a grand potential), we extend the Lagrangian for the two-flavor
NJL model in Minkowski metric by the inclusion of the chemical
potential $\mu$ and obtain
\begin{eqnarray}
&&  L_q=\bar q\Big [\gamma^\nu i\partial_\nu-
m_0+\mu\gamma^0\Big ]q+ G_1\Big [(\bar qq)^2+
(\bar qi\gamma^5\vec\tau q)^2\Big ]+%\nonumber\\
G_2\!\sum_{A=2,5,7}
\Big [\bar q^Ci\gamma^5\tau_2\lambda_{A}q\Big ]
\Big [\bar qi\gamma^5\tau_2\lambda_{A} q^C\Big ],
\label{1}
\end{eqnarray}
where the quark field $q$ is a flavor doublet and a color triplet
as well as a four-component Dirac spinor, $q^C=C\bar q^t$, $\bar
q^C=q^t C$ are charge-conjugated spinors, and
$C=i\gamma^2\gamma^0$ is the charge conjugation matrix ($t$
denotes the transposition operation). Here, the isotopic symmetry
of quarks is implied ($m_0^u=m_0^d=m_0$) and the quark
chemical potential $\mu >0$ is the same for all flavors. Pauli
matrices $\tau^a\; (a=1,2,3)$ act on the flavor indices of quark
fields, while the (antisymmetric) Gell-Mann matrices $\lambda_A$
contract with the color ones; hereafter, the flavor and color
indices are omitted for simplicity. Clearly, the Lagrangian $L_q$
is invariant under transformations by the color $\rm SU(3)_c$ as well
as by the baryon $\rm U(1)_B$ groups. In addition, when the current
quark masses vanish ($m_0=0$), this Lagrangian resumes the (chiral)
$\rm SU(2)_L\times SU(2)_R$ symmetry (chiral transformations affect
flavor indices only).
\footnote{Since $Q=I_3+B/2$, where
$I_3=\tau_3/2$ is the third component of the isospin, and $Q$
and $B$ are electric and baryon charges, respectively, the
Lagrangian (\ref{1}) is invariant under $\rm U(1)_Q$
transformations generated by electric charge, as well.}
Finally, we remark that Lagrangian~(\ref{1}) is $C$-even in the
vacuum, \textit{i.~e.\/}\ it is invariant under the charge
conjugation if $\mu=0$ ($q\to q^C\equiv C\bar q^t$,
$\bar q\to \bar q^C\equiv q^t C$), which is not the case
for dense quark matter where the violation of $C$-parity is induced
by a nonvanishing baryonic chemical potential.

Throughout all our calculations, we assume that the model parameters,
\textit{i.~e.\/} the ultraviolet cutoff $\Lambda$ \footnote{There are
divergent integrals in the model, and a regularization is needed.},
the current quark mass, and the coupling constants, do not change
with $\mu$. Their values are fixed in the vacuum so
that the model can reproduce the experimental value of the pion
mass $M_\pi=140$~MeV, the pion weak-decay constant $F_\pi=92.4$~MeV,
and the value of the chiral quark condensate $\expv{\bar
qq}=-(245~\mbox{MeV})^3$. It has been shown in previous papers that a
convenient set of parameters for the vacuum case is
$G_1=5.86$ GeV$^{-2}$, $\Lambda=618$ MeV, and $m_0=5.67$ MeV, leading
to the constituent quark mass  $350$~MeV. (One can follow,
\textit{e.~g.\/}, the parameter-fixing procedure explained in
\cite{ebvol,bvy03} to obtain values close to these.)  However, the
definition of the constant $G_2$ that describes the interaction of
quarks in the diquark channel is not quite transparent. This constant
is not bounded by some experimental reason, and often, its value is
fixed through a constraint (obtained after the Fierz  transformation)
that connects $G_2$ with constants in the quark-antiquark channels.
Starting from the four-quark vertices provided by the one-gluon
exchange, one  obtains the constraint $G_2=3G_1/4$ which we shall use
in our further calculations.

In principle, all constants in the diquark channels could be fixed if
all interaction constants in the quark-antiquark channels were known.
Unfortunately, only  few of them are available from experiment,
thereby preventing us from a unique definition of the constants in
the diquark channel at the moment. There is still a freedom in fixing
their values. This was the reason  to use some model assumptions,
such as the constraint that follows from the one-gluon exchange
contribution. \footnote{Some constants in the diquark
channel can, however, be extracted from the nucleon mass,
\textit{e.~g.\/} within the Bethe-Salpeter approach
\cite{ebjur,reinh}.}

\subsection{Quark propagator in the case of diquark condensation. Gap
equations}

It is convenient to consider a linearized version of Lagrangian
(\ref{1}) containing auxiliary bosonic fields which is given by the
following form
\begin{eqnarray}
&&L\ds =\bar q\Big [\gamma^\nu i\partial_\nu +\mu\gamma^0
 -\sigma - m_0 -i\gamma^5\pi_a\tau_a\Big ]q
 -\frac{1}{4G_1}\Big [\sigma\sigma+\pi_a\pi_a\Big ]-
 \frac1{4G_2}\Delta^{*}_{A}\Delta_{A}+
 \nonumber\\&& ~~~~~~~~~~~~~~~~~~~~
+\frac{i\Delta^{*}_{A}}{2}[\bar q^Ci\gamma^5\tau_2\lambda_{A} q]
-\frac{i\Delta_{A}}{2}[\bar q i\gamma^5\tau_2\lambda_{A}q^C].
\label{2}
\end{eqnarray}
As usual, the summation over  repeated indices $a=1,2,3$ and
$A,A'=2,5,7$ is implied throughout all our calculations. The
equations of motion for the bosonic fields are
\begin{align}
\sigma (x)
&=-2G_1(\bar qq),
&\Delta_{A}(x)
&=2iG_2(\bar q^Ci\gamma^5\tau_2\lambda_{A}q),\nonumber\\
\pi_a(x)&=-2G_1(\bar qi\gamma^5\tau_a q),
& \Delta^{*}_{A}(x)
&=-2i G_2(\bar qi\gamma^5\tau_2\lambda_{A} q^C).
 \label{3}
\end{align}
Substituting (\ref{3}) into (\ref{2}), one can easily obtain the
initial Lagrangian (\ref{1}). Just in this sense the two theories,
the first one with $L_q$ and the second one with $L$, are equivalent.
It follows from (\ref{3}) that the meson fields $\sigma,\pi_a$
are real, \textit{i.~e.\/} $(\sigma(x))^\dagger=\sigma(x),~~
(\pi_a(x))^\dagger=\pi_a(x)$ (the symbol $\dagger$ stands for the
hermitian conjugation), whereas all diquark fields $\Delta_{A}$ are
complex, $(\Delta_{A}(x))^\dagger=\Delta^{*}_{A}(x)$. Each
$\Delta_{A}$ is an isoscalar  ($\rm SU(2)_L\times SU(2)_R$-singlet).
Moreover, all diquarks are Lorentz scalars and form an
antitriplet ($\bar 3_c$) fundamental representation of the color
$\rm SU(3)_c$ group, while the real scalar $\sigma$ and pseudoscalar
$\pi_a$ fields are color singlets. A nonvanishing value of the scalar
diquark condensate, associated to a nonzero ground-state expectation
value of some diquark field, $\vev{\Delta_{A}}\ne 0$,  breaks
$\rm SU(3)_c$ spontaneously down to $\rm SU(2)_c$ (however, it does
not violate the chiral $\rm SU(2)_L\times SU(2)_R$ symmetry), whereas
a nonzero expectation value of $\vev{\sigma}\ne 0$ at $m_0=0$
indicates that the chiral symmetry is spontaneously broken.
We  assume hereafter that $P$-parity is conserved, \textit{i.~e.\/}\
$\vev{\pi_a(x)}=0$.

Using the Nambu--Gorkov formalism, we put the quark fields and their
charge conjugates together into a bispinor
\hbox{$\Psi=\left(\genfrac{}{}{0pt}{}{q}{q^C}\right)$}, and
Lagrangian (\ref{2}), thereafter, simplifies to
\begin{equation}
\label{4}
L=-\frac{\sigma^2+\vec{\pi}^2}{4G_1}-\frac{\Delta_A\Delta_A^*}{4G_2}+
\frac 12\bar\Psi\left (\begin{array}{cc}
{\cal D}^+, & {\cal K}\\
{\cal K}^*, &{\cal D}^-
\end{array}\right )\Psi,
\end{equation}
where the following notations are adopted
\footnote{In the derivation of (\ref{4}), we used the
well-known relations:
$\partial_\nu^t=-\partial_\nu$, $C\gamma^\nu C^{-1}=-(\gamma^\nu)^t$,
$C\gamma^5C^{-1}=(\gamma^5)^t=\gamma^5$,
$\tau^2\vec\tau\tau^2=-(\vec\tau)^t$,
$\tau^2=\left (\begin{array}{cc}
0~, & -i\\
i~, &0
\end{array}\right )$.}
\begin{eqnarray}
&&{\cal D}^+=i\gamma^\nu\partial_\nu-m_0+\mu\gamma^0-\Sigma,~~~~~~~
{\cal D}^-=i\gamma^\nu\partial_\nu-m_0-\mu\gamma^0-\Sigma^t,
\nonumber\\&&
  \Sigma=\sigma+ i\gamma^5\vec\pi\vec\tau,~~~~\Sigma^t=\sigma+
  i\gamma^5\vec\pi\vec\tau^t,~~~~
  {\cal K}^*=-\Delta^{*}_A\lambda_A\gamma^5\tau^2,\qquad
  {\cal K}=\Delta_A\lambda_A\gamma^5\tau^2.\qquad
\label{5}
\end{eqnarray}
Now, let $\vev{\sigma}\equiv m-m_0\ne 0$ and $\vev{\Delta_{A}}\ne
0$ for some $A$. Without loss of generality, one can always take
advantage of the color symmetry of strong interactions and
rotate the basis of diquark fields so that
$\vev{\Delta_{2}}\equiv\Delta$,
$\vev{\Delta^*_{2}}\equiv\Delta^*$, $\vev{\Delta_{5}}=0$,
$\vev{\Delta_{7}}=0$.
As was shown in previous investigations, $\Delta=0$ in the vacuum and
for small $\mu$, and the quark matter is thereby color symmetric.
Once  $\Delta$ acquires a nontrivial value, the $\rm SU(3)_c$
symmetry is spontaneously broken down to $\rm SU(2)_c$ and the 2SC
phase is formed.
According to the mean-field approximation approach, which we use
here, the mean value of $\Delta$ should be subtracted from the
diquark-quark vertices in (\ref{4}) (or (\ref{2})).
Shifting the fields as follows:
$\sigma (x)\to\sigma (x)+\vev{\sigma}$,
$\Delta^{*}_2(x)\to\Delta^{*}_2(x) +\Delta^{*}$,
$\Delta_2(x)\to\Delta_2(x)+\Delta$,
we absorb the mean values of $\sigma$ and $\Delta$
(with the exception of the quadratic terms) in the inverse quark
propagator $S^{-1}$ and obtain
\begin{equation}
\label{6}
\tilde L=-\frac{\sigma
(m-m_0)}{2G_1}-\frac{\Delta\Delta^*_2+\Delta^*\Delta_2}{4G_2}
-\frac{\sigma^2+\vec{\pi}^2}{4G_1}-\frac{\Delta_A\Delta_A^*}{4G_2}+
\frac12\bar\Psi(S^{-1}+V)\Psi,
\end{equation}
where $\sigma$, $\pi_a$, $\Delta^*_A$ and $\Delta_A$ stand for the
fluctuations around the mean values of mesons and diquarks rather
than the original fields \footnote{One must not be confused with
using the same notations both for the original fields and  their
fluctuations, because the original fields will never appear in
the rest of the paper.}, and
\begin{equation}
\label{7}
S^{-1}=\left (\begin{array}{cc}
i\gamma^\nu\partial_\nu-m+\mu\gamma^0, &
\Delta\lambda_2\gamma^5\tau^2\\
-\Delta^{*}\lambda_2\gamma^5\tau^2,
&i\gamma^\nu\partial_\nu-m-\mu\gamma^0
\end{array}\right ),~~~~~~~~~
V=\left(\begin{array}{ll}
  -\sigma - i\gamma_5\vec\pi \vec\tau, &
  \Delta_A\gamma_5\tau_2\lambda_A\\
  -\Delta_A^*\gamma_5\tau_2\lambda_A, & -\sigma - i\gamma_5\vec\pi
  \vec\tau^t \end{array}\right).
\end{equation}
Now we can perform a series expansion, treating the term $V$,
which contains only fluctuations of meson and diquark fields, as a
perturbation. In the rest of our paper, we shall keep in mind
the Feynman diagram rules for the calculation of two-point field
correlators (Green's functions) in the 2SC phase, however, without
drawing the corresponding graphs. Indeed, the term $S^{-1}$ supplies
us with the quark propagator $S$ in the presence of a diquark
condensate $\vev{qq}$, and the term $V$ is responsible for
quark-meson and quark-diquark vertices. In momentum space, the quark
propagator $S(p)=\left
(\begin{array}{cc}
S_{11}, & S_{12}\\
S_{21}, & S_{22}\end{array}\right )$ (here  $p$ is the
four-momentum of quarks) is represented by a $2\times 2$ matrix, with
respect to the Nambu-Gorkov indices (in addition, it is a $2\times
2$-, $3\times 3$-, as well as $4\times 4$ matrix in the flavor-,
color-, and spinor spaces, correspondingly). After some algebra, one
obtains for $S_{ij}(p)$ ( \cite{bekvy}, \cite{huang}):
\begin{equation}
\label{8}
  S_{ij}(p)=S_{ij}^{\rm rg}(p)\Proj{\rm rg}+S_{ij}^{\rm
  b}(p)\Proj{\rm b},
\end{equation}
where $\Proj{\rm rg}={\rm diag}(1,1,0)$, $\Proj{\rm b}={\rm
diag}(0,0,1)$ are diagonal matrices projecting  onto the
red-green and blue quark components in color space, respectively, and
\begin{eqnarray}
  \label{9}
&&  S_{11}^{\rm rg}=\frac{p_0+E^-}{D_{-}(p_0)}\gamma_0\Ltildeminus+
  \frac{p_0-E^+}{D_{+}(p_0)}\gamma_0\Ltildeplus,~~~~~~~
  S_{12}^{\rm rg}=
  \Delta\gamma_5\tau_2\lambda_2\left(
  \frac{\Ltildeplus}{D_{-}(p_0)}
  +\frac{\Ltildeminus}{D_{+}(p_0)}\right), \nonumber\\
&&  S_{21}^{\rm rg}=-\Delta^*\gamma_5\tau_2\lambda_2\left(
  \frac{\Ltildeplus}{D_{+}(p_0)}
  +\frac{\Ltildeminus}{D_{-}(p_0)}\right),~~~~~~~
  S_{22}^{\rm rg}=\frac{p_0+E^+}{D_{+}(p_0)}\gamma_0\Ltildeminus+
  \frac{p_0-E^-}{D_{-}(p_0)}\gamma_0\Ltildeplus,
\end{eqnarray}
\begin{eqnarray}
\label{10}
  S_{11}^{\rm b}=\frac{\gamma_0\Ltildeminus}{p_0-E^-}+
  \frac{\gamma_0\Ltildeplus}{p_0+E^+},~~~~
  S_{22}^{\rm b}=\frac{\gamma_0\Ltildeminus}{p_0-E^+}+
  \frac{\gamma_0\Ltildeplus}{p_0+E^-},~~~~
  S_{12}^{\rm b}=S_{21}^{\rm b}=0.
\end{eqnarray}
We used in eqs.\ (\ref{9}) and (\ref{10}) the projection operators
\begin{eqnarray}
  \label{11}
  \Lambda_{\pm}(\vec{p})=\frac12\left( 1\pm
				\frac{\gamma_0(\vec{\gamma}\vec{p}+m)}{E}\right),~~~~
				~~~~
  \widetilde{\Lambda}_{\pm}(\vec{p})&=&\frac12\left( 1\pm
				\frac{\gamma_0(\vec{\gamma}\vec{p}-m)}{E}\right)
\end{eqnarray}
to separate the `positive-energy' and `negative-energy' parts of the
quark propagator. Note the following useful properties of these
projectors:
\[
  \gamma_0\widetilde{\Lambda}_{\pm}\gamma_0=\Lambda_{\mp},
  \qquad
  \gamma_5\widetilde{\Lambda}_{\pm}\gamma_5=\Lambda_{\pm}.
\]
Finally, the following notations have been used in eqs.~(\ref{9}) and
(\ref{10})
\begin{equation}
  \label{13}
  D_{+}(p_0)=p_0^2-\Edeltabr{+}^2,\qquad
  D_{-}(p_0)=p_0^2-\Edeltabr{-}^2,~~~~~~
  E_\Delta^\pm=\sqrt{(E^\pm)^2+
  |\Delta|^2},~~~~~~E^\pm=E\pm \mu,
\end{equation}
where $E=\sqrt{\vec{p}^2+m^2}$ is the  dispersion law for free
quarks. The poles of the matrix elements (\ref{9}) and (\ref{10}) of
the quark propagator give the dispersion laws for quarks in the
medium. Thus, we have $E_\Delta^-$ for the energy of red/green quarks
and $E_\Delta^+$ for the energy of red/green antiquarks.
Below, we shall show that $\mu>m$ in the 2SC phase
(see Fig. 1), and $E$ can reach the value of $\mu$.
In this case, in order to create a red/green quark in the 2SC
phase, a minimal amount of energy  (the gap) equal
to $|\Delta|$ at the Fermi level ($E=\mu$) is required. Similarly,
the energy of a blue quark (antiquark) is $E^-$ ($E^+$), hence
$E^-=0$ at $E=\mu$, and there is no energy cost to create a blue
quark, \textit{i.~e.\/} blue quarks are gapless  in the 2SC phase.
(In contrast, to create an antiquark in the 2SC phase, we need the
energy $\sqrt{(m+\mu)^2+|\Delta|^2}$ for red/green antiquarks and the
energy $(m+\mu)$ for blue one.) Given the explicit expression for the
quark propagator $S(p)$, we then calculate two-point correlators of
meson and diquark fluctuations over the ground state in the one-loop
(mean-field) approximation.

The two additional quantities $m$ and $\Delta$ are not free
parameters of Lagrangian (\ref{6}). The constituent quark mass
$m$ is an indicator of the chiral symmetry breaking (if $m_0 =0$, $m$
vanishes when the chiral symmetry is restored). The gap $\Delta$ is
related to the color symmetry breaking/restoration in a similar way.
Both of them are to be found from the requirement that the ground
state expectation values of quantum fluctuations must be zeros.
It is easy to see from (\ref{6}) that in the mean-field (one-loop)
approximation the two conditions $\vev{\sigma}=0$ and
$\vev{\Delta^*_2}=0$ are realized if the following two equations (gap
equations) are true
\begin{equation}\label{14}
  \frac{m-m_0}{2G_1}+\frac{i}{2}\int\frac{d^4q}{(2\pi)^4}
  \Tr[S(q)]=0, ~~~~\frac{\Delta}{4G_2}
  +\frac{i}{2}\int\frac{d^4q}{(2\pi)^4}
  \Tr[S(q)\ \Gamma_{\Delta^*}]=0.
\end{equation}
Here $\Gamma_{\Delta^*}$ stands for the $q$-$\bar q$-$\Delta^*_2$
vertex in the Nambu--Gorkov representation: $ \Gamma_{\Delta^*}=
  \left(\begin{array}[c]{cc}
				  0~~~~ ,& 0\\ -\gamma_5\tau_2\lambda_2 ,& 0
				\end{array}     \right)$.
In (\ref{14}) and in all similar formulas below, the calculation of
the trace includes also a sum over Nambu--Gorkov indices, in addition
to the spinor, color, and flavor ones. (Using the Feynman diagram
terminology, one can say that the first term in each of
eqs.~(\ref{14}) is a tree term whereas the second one represents the
one-loop contribution.) After some trace calculations, we have from
(\ref{14})
\begin{eqnarray}
\frac{m-m_0}{2G_1}&=&4im\!\int\!\frac{d^4q}{(2\pi)^4E}
\left\{\frac{E^+}{q_0^2-(E^+)^2}+\frac{E^-}{q_0^2-(E^-)^2}+
\frac{2E^+}{D_+(q_0)}+
\frac{2E^-}{D_-(q_0)}\right\},\label{140}\\
\frac{\Delta}{4G_2}&=&4i\Delta\!\int\!\frac{d^4q}{(2\pi)^4}
\left\{\frac{1}{D_+(q_0)}+ \frac{1}{D_-(q_0)} \right\}.
\label{141}
\end{eqnarray}

To proceed further, we employ the imaginary-time formalism, used in
theories with finite temperature  and chemical potential in Euclidean
metric to obtain Green's functions. In these theories, at some finite
temperature $T$, the integration over the energy-variable in each
loop is replaced by a sum over Matsubara frequencies. The case of
cold quark matter can be considered simply as the limit $T\to 0$.
One should note here that we are interested in Green's functions in
Minkowski metric rather than in Euclidean, and a continuation from
one metric to the other is needed. Let us consider  Green's functions
as functions of energy only, forgetting, for a moment, about the
three-momentum. At this point, we assume that after the limit $T\to
0$ is taken, the resulting Green's functions are defined in a complex
plane, and we associate the points lying on the imaginary axis with
Euclidean metric. We continue the Green's functions to the real axis
and consider the thus obtained new functions as being defined
in Minkowski metric. Clearly, one should impose the constraint that
such a continuation must reproduce the result obtained from quantum
field theory in Minkowski metric when we put $\mu=0$.

To calculate the integrals in (\ref{140}) and (\ref{141}), we
replace the integration over $q_0$ by a sum over  Matsubara
frequencies, $\omega_n =(2n+1)T$,\, $n=0, \pm 1, \pm 2,..$, followed
by the limit $T\to0$:
\begin{equation}
\label{15}
\int\frac{d^4q}{(2\pi)^4}f(q_0,\vec q)\longrightarrow
i\lim_{T\to 0}T\sum_n\int\frac{d^3q}{(2\pi)^3}f(i\omega_n,\vec q).
\end{equation}
Applying rule (\ref{15}) to eqs.\ (\ref{140}) and (\ref{141}),
we obtain  the gap equations for  $m$ and $\Delta$ in the
cold quark matter ($T=0$) (for more explanations, see Appendix A):
\begin{eqnarray}
\frac{m-m_0}{2G_1}&=&4m\int\frac{d^3q}{(2\pi)^3E}
\Big\{\theta(E^-)+\frac{E^+}{E_\Delta^+}+\frac{E^-}{E_\Delta^-}\Big\}
,
\label{16}\\
\frac{\Delta}{4G_2}&=&2\Delta\int \frac{d^3q}{(2\pi)^3}
\Big\{\frac{1}{E_\Delta^+}+
\frac{1}{E_\Delta^-}\Big\}\equiv\Delta I_\Delta.
\label{17}
\end{eqnarray}
Since the integrals in the right hand sides of these equations are
ultraviolet divergent, we regularize them and the other divergent
integrals below by implementing a three-dimensional cutoff $\Lambda$.

The system of eqs.\ (\ref{16}) and (\ref{17}) has two different
solutions. As we have already discussed, the first one (with
$\Delta=0$) corresponds to the $\rm SU(3)_c$-symmetric phase of the
model (normal phase), the second one (with $\Delta\ne 0$) to the 2SC
phase. As usual, solutions of these equations give  local extrema of
the thermodynamic potential $\Omega (m,\Delta)$
\footnote{An expression for the thermodynamic potential for a system
of free  fermions can be found elsewhere; see, \textit{e.~g.\/},
\cite{bekvy}.}, so one should also check which of them corresponds to
the absolute minimum of $\Omega$. Having found the solution
corresponding to the stable state of quark matter (the absolute
minimum of $\Omega$),
we obtained the behavior of the gaps $m$ and $\Delta$ \textit{vs.\/}
chemical potential (see Fig.~1). The region $\mu <\mu_c = 350$~MeV
is the domain of color symmetric quark matter because $\Omega$  in
this case is minimized by $m\ne 0$ and $\Delta =0$. For  $\mu>
\mu_c$, the color symmetric phase becomes unstable because a solution
with $\Delta =0$ does not  minimize  $\Omega$. Here, the solution
with $m\ne 0$ and $\Delta\ne 0$, corresponding to the 2SC phase,
gives the global minimum  of $\Omega$, and thereby the color
superconducting phase is favored (note that in the 2SC phase $\mu
>m$, whereas $\mu <m$ in the color symmetric one).
The transition between these two phases is of the first-order, which
is characterized by a discontinuity in the behavior of
$m$ \textit{vs.\/} $\mu$ (see Fig.~1).
\begin{figure}
\begin{center}
\includegraphics[scale=0.5]{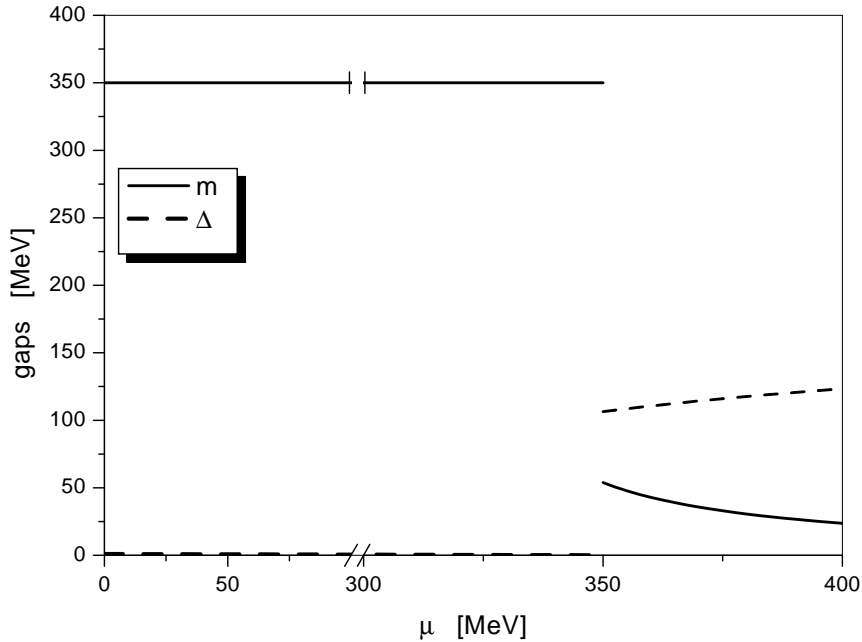}
\end{center}
\caption{The constituent quark mass $m$ (solid line) and the
color gap $\Delta$ (dashed line) as  functions of the chemical
potential $\mu$. } \label{MDeltamu1}
\end{figure}

\section{Mesons and diquarks in dense quark matter}

We are now interested in the investigation of the modification of
meson and diquark masses in dense and cold quark matter with the
color symmetry broken because of the 2SC diquark condensation. In the
vacuum ($T=0$, $\mu=0$), particle masses are obtained from propagator
poles, or alternatively, from zeros of the one-particle irreducible
(1PI) two-point Green's functions. Since the Lorentz invariance is
preserved in the vacuum, these functions in (Minkowski) momentum
space depend on $p^2=p_0^2-\vec p^2$ only. The squared mass of the
particle is thus equal to the value of $p^2$ where the corresponding
two-point 1PI correlators (Green's function) vanishes. Insofar as
$p^2$ is a Lorentz invariant, one can, for simplicity, choose the
rest frame ($\vec{p}=0$), put $p^2=p_0^2$ and then consider these
correlators as functions of $p_0$ alone. On the contrary, in a dense
medium, the Lorentz invariance is broken, so the two-point Green's
functions in momentum space should be treated  as functions of two
variables: $p_0$ and $\vec{p}^2$ (the rotational symmetry is
presumably conserved). The zeros of 1PI two-point Green's functions
in the $p_0$ plane will determine the particle and antiparticle
dispersion laws, \textit{i.~e.\/}\ the relations between their
energy and three-momenta. In this case, the scalar particle mass
is defined as the value of the particle energy at $\vec p=0$ (see,
\textit{e.~g.\/}~\cite{ruivo}). Recently,  the Bethe-Salpeter
equation approach has been used to obtain diquark masses in the 2SC
phase of cold dense QCD at asymptotically large values of the
chemical potential \cite{mirsh}. There, the mass of the diquark was
defined as the energy of a bound state of two virtual quarks in
the center of mass frame, \textit{i.~e.\/} in the rest frame for the
whole diquark. Let us note here that just this quantity is measured
on the lattice, where it is given by the exponential fall-off of the
particle propagator at large Euclidean time (see, \textit{e.~g.\/},
\cite{hands}).%
\footnote{In the last case
the lattice calculations were performed in some QCD-like theories
(QCD with two colors, etc), where the fermion determinant is
positive even at nonzero $\mu$. Evidently, this mass definition is
borrowed from particle physics. Note that in condensed
matter physics usually the term "energy gap" is used for this
quantity (see, \textit{e.~g.\/}, \cite{miransky}, where both terms,
"mass" and  "energy gap", are used for the rest frame energy
of scalar particles). }
Similarly as in \cite{ruivo,ratti} and in numerous other papers,
we shall denote, in the following, the rest frame energy of a
composite scalar particle (meson or diquark), moving in a dense
medium, as ``mass''. (In general, the values of the mass  depend
on the chemical potential.)

Any 1PI Green's function can be found from the effective action
$S_{\rm eff}$, which up to second order in boson fields has the form
\[
S_{\rm eff}=\frac 12\sum_{X,Y}\int d^4xd^4y
X(x)\Pi_{XY}(x,y)Y(y)+\cdots,
\]
where $X,Y=\pi_a,\,\sigma,\, \Delta^*_A,\,\Delta_B$, and
$\Pi_{XY}(x,y)$ is the coordinate  representation of the 1PI Green's
function for the fields $X,Y$. Instead of employing functional
integration to get $S_{\rm eff}$ and then $\Pi_{XY}$, we shall use
implicitly Feynman diagrams. Starting from the Lagrangian (\ref{6}),
one can expand the resulting effective action $S_{\rm eff}$ in a
power series of meson and diquark fluctuations. Keeping there only
the second order contributions, we immediately obtain the 1PI
two-point correlators in the one-loop approximation. By considering
this functions as functions of $p_0$ at zero three-momentum
($\vec{p}=0$), we shall next analyze their zeros which,
as explained above, will give the masses of resonances.

\subsection{Pion mass}

Let us begin with the calculation of the pion mass. In the  momentum
representation, the 1PI two-point function $\Pi_{\pi_a\pi_b}(P)$ for
the pion has the following form (all calculations are performed
in the rest frame, $P=(p_0,0,0,0)$ )
\begin{equation}
  \label{18}
  \Pi_{\pi_a\pi_b}(P)=-\frac{\delta^{ab}}{2G_1}+ \frac{i}{2}\int
  \frac{d^4q}{(2\pi)^4}
  \Tr\left[S(P+q)\Gamma_\pi^a S(q)\Gamma_\pi^b\right].
\end{equation}
Here, the vertex of the pion-quark interaction is given by
the 2$\times$2 matrix $\Gamma_\pi^a=\left(\begin{array}[c]{ll}
i\gamma_5\tau_a~ & 0\\
0~&i\gamma_5\tau_a^t\end{array}\right)$. The first term in the right
hand side of (\ref{18}) is the tree contribution from Lagrangian
(\ref{6}), the second term arises from the one-loop  diagram
with two pion legs.  After intermediate trace calculations, this
function takes the form
\begin{eqnarray}
&&\Pi_{\pi_a\pi_b}(P)=-\frac{\delta_{ab}}{2G_1}
+16i\delta_{ab}\int\frac{d^4q}{(2\pi)^4}
\frac{q_0(p_0+q_0)-E^+E^--\Delta^2}{D_-(q_0)D_+
(p_0+q_0)}+\nonumber\\
&&\qquad+8i\delta_{ab}\int\frac{d^4q}{(2\pi)^4}
\frac{q_0(p_0+q_0)-E^+E^-}{
[(p_0+q_0)^2-(E^+)^2][(q_0)^2-(E^-)^2]}.
\label{19}
\end{eqnarray}
Implementing the imaginary-time formalism, as described in the end
of the previous section (see also Appendix A for details), we reduce
eq.\ (\ref{19}) to  three-dimensional integration in momentum space
\begin{eqnarray}
&&\Pi_{\pi_a\pi_b}(P)=-\frac{\delta_{ab}}{2G_1}
+8\delta_{ab}\int\frac{d^3q}{(2\pi)^3}
\frac{E_\Delta^+E_\Delta^-+E^+E^-+\Delta^2}{E_\Delta^+E_\Delta^-}
\frac{E_\Delta^++E_\Delta^-}{(E_\Delta^++E_\Delta^-)^2-p_0^2}+
\nonumber\\
&&\qquad+16\delta_{ab}\int\frac{d^3q}{(2\pi)^3}
\frac{\theta (E-\mu)E}{4E^2-p_0^2}\equiv
\delta_{ab}\Pi_{\pi\pi}(p_0).
\label{20}
\end{eqnarray}
Then, up to a sign, the pion unnormalized propagator equals
$(\Pi_{\pi_a\pi_b}(P))^{-1}$, whose pole in $p_0$ is given by the
zero of the function $\Pi_{\pi\pi}(p_0)$ from (\ref{20}).
We have searched for the roots of the equation $\Pi_{\pi\pi}(p_0)=0$
numerically; the results for the pion mass $M_\pi$ at various $\mu$
are plotted in Fig.~2. A similar behavior of the pion mass in the
Cooper pairing phase of a dense quark matter was found in
\cite{ratti} in the framework of a two-colored NJL model.
\begin{figure}
  \centering
  \includegraphics[width=14cm]{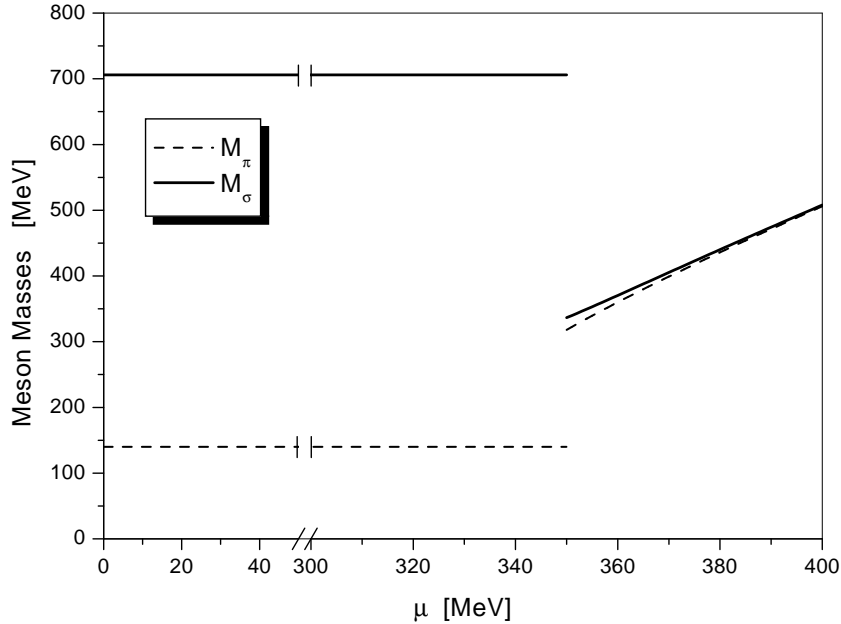}
  \caption{The masses of the $\sigma$-meson (solid line) and pion
  (dashed line) as functions of $\mu$.}\label{plot:mesonmasses}
\end{figure}

In the color symmetric phase, the pion mass is below the
threshold for the pion decay to a quark-antiquark pair, and the pion,
therefore, is almost stable (only electroweak decay channels are
allowed). Moreover, we have found that in the 2SC phase the pion is
also an almost stable particle. This conclusion is supported by the
following
arguments. On the one hand, the pion cannot decay into a red/green
quark-antiquark pair, since $M_\pi$ is less than the minimal energy
$E_{\text{min}}= \sqrt{(\mu-m)^2+|\Delta|^2}+ \sqrt{(\mu+m)^2+
|\Delta|^2}$ that is necessary to create these particles (evidently,
$E_{\text{min}}$ is the value of $E_\Delta^-+E_\Delta^+$ (see eq.
(\ref{13}) and the text after it) at $E=m$).
On the other hand, the decay of a pion into a pair of blue
quark-antiquark is also forbidden: i) Since all states with $E<\mu$
are occupied by quarks from the medium, the pion cannot decay into a
blue quark with such values of $E$ because of the Pauli blocking,
i.e. the analogy to the Mott effect. ii) In the region $E\ge
\mu$, the energy of a blue quark-antiquark pair $(E^-+E^+)$ takes
its least value $E_{\text{min}}=2\mu$ at $E=\mu$, so the pion decay
is again impossible, since $M_\pi<2\mu$.  As a consequence, the pion
is almost stable in the 2SC phase, too.

\subsection{Mixing of $\sigma$-$\Delta_2$ in the 2SC phase. Scalar
meson mass}
\label{3.B}

An investigation of the 1PI Green's functions $\Pi_{\sigma X}(P)$
with $P=(p_0,0,0,0)$ for the fields $X= \Delta^*_A, \Delta_B$ shows
that in the 2SC phase the $\sigma$-meson is mixed with the $\Delta_2$
diquark. (At $\mu>\mu_c$, such a mixing occurs in the NJL model with
two-colored quarks \cite{ratti}, too. Moreover, as our preliminary
results show, the mixing is present even if the condition of color
neutrality of the 2SC phase is imposed.) The mass of the
$\sigma$-meson in this case is given by the solution of the equation
${\rm det}(\Pi)=0$ where $\Pi (P)$ is the 3$\times$3 matrix
\begin{eqnarray}
\Pi (P)=\left(\begin{array}[c]{lll}
\Pi_{\sigma\sigma}(P)~, & \Pi_{\sigma\Delta_2}(P),& \Pi_{\sigma
\Delta^*_2}(P)\\
\Pi_{\Delta_2\sigma}(P),&\Pi_{\Delta_2\Delta_2}(P),&\Pi_{\Delta_2
\Delta^*_2}(P)\\
\Pi_{\Delta^*_2\sigma}(P),&\Pi_{\Delta^*_2\Delta_2}(P),&
\Pi_{\Delta^*_2\Delta^*_2}(P)\end{array}\right).
\label{21}
\end{eqnarray}
(Up to a sign, $\Pi (P)$ is the inverse propagator matrix for
the $\sigma$-meson and $\Delta^*_2$, $\Delta_2$ diquarks.)
After tedious but straightforward calculations, similar to those
in the pion case, we get
\begin{eqnarray}
\Pi_{\sigma\sigma} (P)&=&-\frac 1{2G_1}+
16\Delta^2m^2\int\frac{d^3q}{(2\pi)^3E^2}\,
\left\{\frac{1}{E_\Delta^+[4(E_\Delta^+)^2-p_0^2]}
+\frac{1}{E_\Delta^-[4(E_\Delta^-)^2-p_0^2]}\right\}+
\nonumber\\
&+&8\int\frac{d^3q}{(2\pi)^3}\,\frac{\vec q^2}{E^2}\,
\frac{E_\Delta^+E_\Delta^-+E^+E^-+\Delta^2}{E_\Delta^+E_\Delta^-}
\,\frac{E_\Delta^++E_\Delta^-}{(E_\Delta^++E_\Delta^-)^2-p_0^2}
+16\int\frac{d^3q}{(2\pi)^3}\,\frac{\vec q^2}{E}\,
\frac{\theta(E-\mu)}{4E^2-p_0^2},
\label{22}\\
\Pi_{\sigma\Delta_2}(P)&=&\Pi_{\Delta^*_2\sigma}(P)=
\Pi_{\sigma\Delta^*_2}(-P)=\Pi_{\Delta_2\sigma}(-P)=\nonumber\\
&=&4m\Delta\int\frac{d^3q}{(2\pi)^3}
\left\{\frac{2E^++p_0}{EE_\Delta^+[p_0^2-4(E_\Delta^+)^2]}
+\frac{2E^--p_0}{EE_\Delta^-[p_0^2-4(E_\Delta^-)^2]}\right\},
\label{23}\\
\Pi_{\Delta_2\Delta_2}(P)&=&\Pi_{\Delta^*_2\Delta^*_2}(P)=
4\Delta^2I_0(p_0^2),~~~~~~~\Pi_{\Delta_2\Delta^*_2}(P)=
\Pi_{\Delta^*_2\Delta_2}(-P)=\nonumber\\
&=&-\frac{1}{4G_2}+I_\Delta+(4\Delta^2-2p_0^2)I_0(p_0^2)+
4p_0I_1(p_0^2),
\label{24}
\end{eqnarray}
where $I_\Delta$ is given by (\ref{17}), $I_0(p_0^2)=A_++A_-$,
$I_1(p_0^2)=B_+-B_-$, and
\begin{eqnarray}
A_+=\int\frac{d^3q}{(2\pi)^3}
\frac{1}{E_\Delta^+[p_0^2-4(E_\Delta^+)^2]},~~~&&~~~A_-=\int
\frac{d^3q}{(2\pi)^3}
\frac{1}{E_\Delta^-[p_0^2-4(E_\Delta^-)^2]},\label{25}\\
B_+=\int\frac{d^3q}{(2\pi)^3}
\frac{E^+}{E_\Delta^+[p_0^2-4(E_\Delta^+)^2]},~~~&&~~~
B_-=\int\frac{d^3q}{(2\pi)^3}
\frac{E^-}{E_\Delta^-[p_0^2-4(E_\Delta^-)^2]}.\label{26}
\end{eqnarray}
(The formulas (\ref{22})--(\ref{24}) are valid both in the 2SC
($\Delta\ne 0$) and in the color symmetric ($\Delta =0$) phases.
In the 2SC phase, the 1PI Green's functions
$\Pi_{\Delta_2\Delta^*_2}(P)$, $\Pi_{\Delta^*_2
\Delta_2}(P)$ (\ref{24})  get a more simple form if one uses
the identity $1=4G_2 I_\Delta$ following from the gap
equations (see eq.\ (\ref{17})).) In the general case ($m\ne 0$,
$\Delta\ne 0$), the equation ${\rm det}(\Pi)=0$ has a rather
complicated form. Fortunately, in the color symmetric phase with
$\Delta =0$, \textit{i.~e.\/}\ at $\mu<\mu_c$,
the nondiagonal terms $\Pi_{\sigma X}(P)$ ($X=\Delta^*_2, \Delta_2$)
in (\ref{21}), which are responsible for the mixing of the
$\sigma$-meson and the diquark, vanish because they are proportional
to $\Delta$. Therefore, the $\sigma$-meson mass decouples from the
diquark spectrum and is found from the equation
\begin{eqnarray}
\Pi_{\sigma\sigma} (P)=0.
\label{27}
\end{eqnarray}
On the other hand, in the 2SC phase (see Fig.~1), the constituent
quark mass $m$ is small (or even equal to zero if $m_0=0$) in the 2SC
phase (see Fig.~1), so one can ignore the nondiagonal elements
$\Pi_{\sigma \Delta_2}(P)$, $\Pi_{\sigma\Delta^*_2}(P)$ in $\Pi$
because  they are negligibly small (as proved \textit{a posteriori}
by numerical computations), and the $\sigma$-meson mass $M_\sigma$
is again found from eq.\ (\ref{27}). The numerical solution of eq.\
(\ref{27}) is presented in Fig.~2, where one can see that both
$\sigma$- and $\pi$-meson masses are increasing functions of $\mu$ in
the 2SC phase.
\footnote{Note that the term in (\ref{22}) that is proportional to
$\Delta^2m^2$ is comparable or even less than
nondiagonal elements $\Pi_{\sigma X}(P)$ ($X=\Delta^*_2, \Delta_2$).
So it was neglected in numerical analysis of (\ref{27}).}
At the same time, the difference between $M_\sigma$
and $M_\pi$ decreases with $\mu$; $\delta M=M_\sigma-M_\pi$ becomes
negligible at sufficiently high $\mu$, which is understood as an
evidence of the chiral symmetry restoration. The decrease of the
dynamical quark mass $m$ at large $\mu$ (see Fig.~1) is also in
accordance with this conclusion.

Finally, we would like to note that the $\sigma$-meson is almost
stable in
the 2SC phase for the same reason which was explained in the last
paragraph of the previous section for the case of the pion.

\subsection{Diquark masses}

In dense quark matter (at nonzero (baryon) chemical potential), the
symmetry of the Lagrangian under charge conjugation is violated by
the chemical potential. As a consequence, the mass spectrum of
diquarks can split, and  diquarks will differ from antidiquarks not
only by charge but also by mass.

\subsubsection{Diquark masses in the color symmetric phase ($\Delta
=0$)}

In the color symmetric phase  ($\mu<\mu_c =350$~MeV), the ground
state of the quark matter is described by $\Delta =0$, and there is
no mixing between diquarks in the one-loop approximation.
Therefore, one can, \textit{e.~g.\/}, consider the propagator of
$\Delta^*_2, \Delta_2$ alone. It follows from eq.\ (\ref{24}) that
 $\Pi_{\Delta_2\Delta_2}(P)=$
$\Pi_{\Delta^*_2\Delta^*_2}(P)=0$ at $\Delta =0$, and one needs only
\begin{eqnarray}
&&\Pi_{\Delta_2\Delta^*_2}(P)=\Pi_{\Delta^*_2\Delta_2}(-P)=-\frac
1{4G_2}-16
\int\frac{d^3q}{(2\pi)^3}\frac{E}{(p_0+2\mu)^2-4E^2}\equiv
-\frac 1{4G_2}+F(\epsilon ).\label{28}
\end{eqnarray}
Here, $P=(p_0,0,0,0)$, $\epsilon =(p_0+2\mu)^2$. (In obtaining
(\ref{28}), we have used the relation $\mu<m$, \textit{i.~e.\/}\
$E^-> 0$, which is true for the  color symmetric phase.)
Then, the 2$\times$2 inverse propagator matrix
$\mathcal{G}^{-1}_2(P)$ in the $\Delta^*_2,\Delta_2$-sector of the
NJL model has the form
\begin{equation}
  \label{29}
\mathcal{G}^{-1}_2(P)=-  \left(\begin{array}[c]{cc}
			  0~~,& \Pi_{\Delta_2\Delta^*_2}(P)\\
			  \Pi_{\Delta^*_2\Delta_2}(P),&~0\end{array}\right).
\end{equation}
Clearly, the mass spectrum is determined by the equation ${\rm
det}(\mathcal{G}^{-1}_2)= \Pi_{\Delta_2\Delta^*_2}(P)
\Pi_{\Delta^*_2\Delta_2}(P) =0$, or by zeros of (\ref{28}),
where the function $F(\epsilon)$ is analytical in
the whole complex $\epsilon$-plane, except for the cut
$4m^2<\epsilon$ along the real axis. (In general, the function
$F(\epsilon)$ is defined on a complex Riemann surface which is
described by several sheets. However, a direct numerical computation
based on eq.\ (\ref{28}) gives its values in the first sheet only.
To find a value on the rest of the Riemann surface, a special
procedure of continuation is needed.) The numerical analysis of
(\ref{28}) in the first Riemann sheet shows that the equation
$\Pi_{\Delta_2\Delta^*_2}(P)=\Pi_{\Delta^*_2\Delta_2}(-P)=0$ has
a root ($\epsilon_0$) on the real axis ($0<\epsilon_0<4m^2$),
providing us with the following massive diquark modes
\begin{equation}
  \label{30}
M_\Delta=1.998 m-2\mu,~~~~~M_{\Delta^*}=1.998 m+2\mu.
\end{equation}
We relate   $M_\Delta$ in (\ref{30}) to the mass of the diquark with
the baryon number $B=2/3$ and $M_{\Delta^*}$ to the mass of
the antidiquark with $B=-2/3$. (Qualitatively, a similar
behavior of diquark and antidiquark masses \textit{vs.\/} $\mu$ was
obtained in \cite{ratti} in the NJL model for two-colored quarks.)
It follows from (\ref{30}) that in the vacuum ($\mu=0$) the
diquark/antidiquark mass is $\sim 2m$. Clearly, in the color
symmetric phase at $\mu\ne 0$ both quantities $M_\Delta$ and
$M_{\Delta^*}$ from (\ref{30}) are nothing else but the rest frame
excitation energies for a diquark and antidiquark, respectively.

In the color symmetric phase for $\mu <m$ the diquarks and
antidiquarks are stable. Indeed, the diquark mass
$M_\Delta$ is smaller here than the energy $2(m-\mu)$
necessary to create a pair of free quarks. Finally,  due to the
underling color $\rm SU(3)_c$ symmetry, the previous statement is
valid also for $\Delta^*_5, \Delta_5$ and $\Delta^*_7, \Delta_7$. As
a result, we have a color antitriplet of diquarks  with
the mass $M_\Delta$ (\ref{30})  as well as a color triplet of
antidiquarks with the mass $M_{\Delta^*}$. The results of numerical
computations are presented in Fig.~3: the solid line shows
the behavior of the antidiquark triplet mass $M_{\Delta^*}$
in the region of $\mu<\mu_c =350$~MeV whereas the dashed line
corresponds to the antitriplet.

In our analysis, we have used the constraint $G_2=3G_1/4$, thereby
fixing the constant $G_2$ through $G_1$. It is useful, however, to
discuss now the influence of  $G_2$  on diquark masses. Indeed, it
is clear from (\ref{28}) that the root $\epsilon_0$  lies inside the
interval $0<\epsilon_0<4m^2$ only if $G_2^*<G_2<G_2^{**}$, where
$G_2^*$ and $G_2^{**}$ are defined by
\begin{eqnarray}
G_2^* &\equiv&\frac {1}{4F(4m^2)}= \frac {\pi^2}{4\left
[\Lambda\sqrt{m^2+\Lambda^2}+m^2\ln((
\Lambda+\sqrt{m^2+\Lambda^2})/m)\right ]},\nonumber\\
G_2^{**}&\equiv&\frac {1}{4F(0)}=\frac {\pi^2}{4\left
[\Lambda\sqrt{m^2+\Lambda^2}-m^2\ln((
\Lambda+\sqrt{m^2+\Lambda^2})/m)\right ]}=\frac{3mG_1}{2(m-m_0)}.
\label{gamma4}
\end{eqnarray}
In this case, there are stable diquarks and antidiquarks in the
color symmetric phase. The behavior of their masses qualitatively
resembles that given by eqs.~(\ref{30}). For a rather weak
interaction in the diquark channel ($G_2<G_2^*$),  $\epsilon_0$ runs
onto the second Riemann sheet, and unstable diquark modes
(resonances) appear. Unlike this, a sufficiently strong interaction
in the diquark channel ($G_2>G_2^{**}$) pushes $\epsilon_0$ towards
the negative semi-axis, \textit{i.~e.\/}\ $(p_0+2\mu)^2<0$. The
latter indicates a tachyon singularity in the diquark propagator,
evidencing that the color symmetric ground state is not stable.
Indeed, as it has been shown in \cite{ek}, at a very large  $G_2$
the color symmetry is spontaneously broken even at a vanishing
chemical potential.

\begin{figure}
  \centering
  \includegraphics[width=14cm]{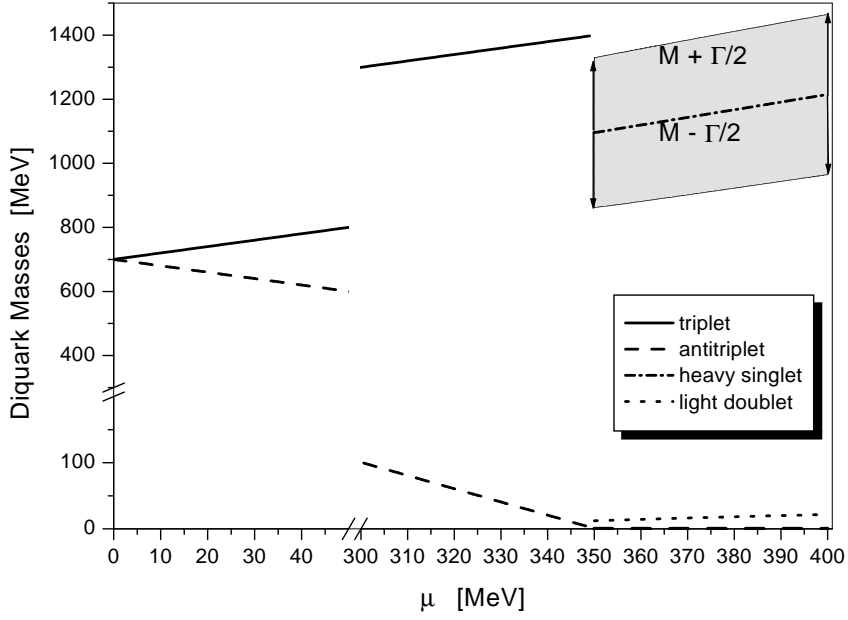}
  \caption{The masses of diquarks. At $\mu<\mu_c=350$ MeV, six
  diquark states are splitted into a (color)triplet of heavy states
  with mass $M_{\Delta^*}$ (solid line) and an anti-triplet (dashed
  line) of light states with mass $M_\Delta$. In the 2SC phase
($\mu>\mu_c$), one observes 3 massless diquarks (dashed line): a
doublet of light diquarks with the mass
$M_{\mbox{\tiny light}}$ (dotted line) and a heavy
singlet state with the mass $M$ (dash-dotted line). The shaded
rectangular displays the width of the heavy singlet resonance; its
upper border is half-width higher than the mass and the bottom
border is half-width lower.}\label{plot:diquarkmasses}
\end{figure}

\subsubsection{Diquark masses in the 2SC phase ($\Delta\ne 0$)}

Let us now focus  on the masses of $\Delta^*_2, \Delta_2$-fields.
As we have already shown in Section \ref{3.B}, these diquarks are
mixed with the $\sigma$-meson in the 2SC phase because of the
nonvanishing terms  $\Pi_{\sigma \Delta_2}(P)$ and
$\Pi_{\sigma\Delta^*_2}(P)$ in the matrix $\Pi (P)$ (\ref{21}).
However, keeping in mind that the constituent quark mass
$m$ is small in the color superconducting phase, one can ignore this
mixing. The problem becomes, thereby, drastically simplified, and
one just has to calculate the determinant of
\begin{equation}
  \label{31}
\mathcal{G}^{-1}_2(P)=-  \left(\begin{array}[c]{cc}
		   \Pi_{\Delta_2\Delta_2}(P),&
		   \Pi_{\Delta_2\Delta^*_2}(P)\\
	  \Pi_{\Delta^*_2\Delta_2}(P),&
	  \Pi_{\Delta^*_2\Delta^*_2}(P)
				\end{array}\right)
\end{equation}
and equate it to zero. The resulting equation will determine the
masses of $\Delta^*_2, \Delta_2$. Taking the expressions for the
matrix elements in (\ref{31}) from eqs.~(\ref{23}) and
(\ref{24}) and using the relation $1=4G_2I_\Delta$  (valid in the
2SC phase only), we get the mass equation
\begin{equation}
  \label{32}
{\rm det}(\mathcal{G}^{-1}_2)\equiv
4p_0^2\big\{(p_0^2-4\Delta^2 )I_0^2(p_0^2)-4I_1^2(p_0^2)\big\}=0.
\end{equation}
It has the apparent solution $p_0^2=0$ corresponding to a
Nambu--Goldstone (NG) boson.%
\footnote{The equation ${\rm det}(\Pi
(p_0))=0$ where $\Pi (p_0)$ is given by (\ref{21}) also has a
NG-solution ($p_0^2=0$) in the 2SC phase. Indeed, one can easily
see that the elements of the second and third columns of $\Pi (p_0)$
are equal at $p_0=0$, and the determinant of this matrix is thereby
equal to zero at $p_0=0$. Since ${\rm det}(\Pi (p_0))$ is an even
function of $p_0$, one can conclude that this solution is doubly
degenerated.}
The second solution of (\ref{32}) exists on the second Riemann
sheet for $p_0^2$ only (see Appendix B).

Near a zero, the determinant ${\rm det}(\mathcal{G}^{-1}_2)$
can be approximated by
\begin{equation}
 {\rm det}(\mathcal{G}^{-1}_2)\sim p_0^2-M^2+i M\Gamma.
\end{equation}
Here, $M$ is the mass of the resonance, and $\Gamma$ is its width.
Let $\tilde{p}_0$ be a root of the equation
\begin{equation}\label{masseq}
 \tilde{p}_0^2-M^2+i M\Gamma=0,
\end{equation}
the mass and width are then given by
\begin{equation}
\label{33a}
 M=\sqrt{\mbox{Re}\,\tilde{p}_0^2 },\qquad
\Gamma=-\frac{\mbox{Im}\,\tilde{p}_0^2}{M}.
\end{equation}
For a small width, we can write for the root of eq.~(\ref{masseq})
 \begin{equation}
  \label{33}
\tilde{p}_0\approx M-i\frac{\Gamma}{2}.
\end{equation}
The appearance of the imaginary part in (\ref{33}) is a consequence
of the fact that this diquark is a resonance in the 2SC phase
and can decay into free quarks. Our numerical estimates for $M$ and
$\Gamma$ at various $\mu$ are plotted in Fig.~3. The dot-dashed line
corresponds to $M$, while the width is given by the hight of the
shaded block, half-width up and half-width down.

We would like to point out here that both the resonance and the
above mentioned NG-boson are color singlets with respect to $\rm
SU(2)_c$. Since the obtained mass $M$ is much greater than even twice
the energy $\sqrt{(m+\mu)^2+|\Delta|^2}$ necessary for creating an
antiquark in the 2SC phase, there is no wonder that this diquark
mode is unstable, unlike the pion and $\sigma$, which are stable due
to the Mott effect.

A detailed investigation of the diquark masses in the $\Delta^*_5,
\Delta_5$- and $\Delta^*_7, \Delta_7$ sectors has  already been done
in our previuos paper \cite{bekvy}. It was found there that in each
of these sectors there was an NG-boson as well as a light diquark
excitation with the same mass $M_{\mathrm{light}}$  proportional to
the color-8 charge density of the quark matter in the 2SC  ground
state. So, we can conclude that, in total, there is an abnormal
number of three NG-bosons in the theory, instead of expected five
(Since $\rm SU(3)_c$ is spontaneously broken down to $\rm SU(2)_c$,
five group generators affect the diquark condensate, and therefore,
five NG bosons should appear, according to the Goldstone theorem).
This entails absence of  baryon superfluidity in the 2SC phase (see
also the discussions in \cite{miransky,ms}, where the effect of
NG-boson deficiency was observed in other relativistic models with
broken Lorentz invariance). The light diquarks form a stable $\rm
SU(2)_c$-doublet with the mass $M_{\mathrm{light}}$ whose dependence
on $\mu$ is also shown in Fig.~3 (dotted line).

\section{Summary and discussion}

In the present paper, we have investigated the mass spectrum of meson
and diquark excitations in cold dense quark matter. We started from
a low-energy effective model of the Nambu--Jona-Lasinio type for
quarks of two flavors including a single quark chemical potential,
for simplicity. Despite of the lack of confinement, this model quite
satisfactorily describes the masses and dynamics of light mesons in
normal quark matter (at rather small values of chemical potential).
Since the investigation of color superconductivity became popular
nowadays, NJL models have also been widely used to explore, in
particular, the quark matter phase diagram for intermediate
densities, \textit{i.~e.\/} under conditions where all other
approaches fail.

Using the one-loop approximation, we calculated two-point
correlators of mesons and showed that the masses of $\pi$- and
$\sigma$-mesons grow with the quark chemical potential in the 2SC
phase (see Fig.~2). The mass difference between them vanishes at
asymptotically large $\mu$, in accordance with chiral symmetry
restoration. Moreover, these mesons are stable in the 2SC phase due
to the Mott effect. As far as we know, the properties of $\pi$- and
$\sigma$-mesons in the 2SC phase have not been discussed in the
literature before.

In the diquark sector, the situation is more involved in the 2SC
phase. Indeed, when the color $\rm SU(3)_c$-symmetry is spontaneously
broken down to $\rm SU(2)_c$, one naturally expects five (massless)
NG bosons  to appear. Unlike this, one finds only three massless
bosonic excitations \cite{bekvy}: a color singlet and a color
doublet (due to the residual $\rm SU(2)_c$ symmetry). In spite of the
abnormal number of NG-bosons (notice that each member of the doublet
has a quadratic dependence of its energy on three-momentum when it is
almost at rest), there is no contradiction with the NG-bosons
counting \cite{nc}. Apart from this, there are also two light and one
heavy diquark modes (see~Fig.~3). The first two are stable, whereas
the last one is a resonance with finite width, and its 1PI Green's
function possesses a zero in the second Riemann sheet for the energy
variable.

We have also found that the antidiquark masses exceed those of the
diquarks in the color symmetric phase  (for $\mu <\mu_c=350$~MeV).
This splitting of the masses is explained by the violation of
$C$-parity (charge conjugation) in the presence of a chemical
potential. In contrast, at $\mu =0$ the model is $C$-invariant and
all diquarks and antidiquarks have the same mass which is
slightly lower than two dynamical quark masses,  $\sim$700~MeV.
Our result for the diquark mass in the vacuum ($\mu=0$) is in
agreement with  \cite{maris}, where the value as large as
$\sim800$~MeV was claimed to follow from QCD via solving a
Bethe--Salpeter equation in the rainbow-ladder approximation.

Of course, all observable particles render themselves as
colorless objects in the hadronic phase, and the diquarks are
expected to be confined, as they are not $\rm SU(3)_c$ color
singlets. Nevertheless, one may look at our and other related results
on diquark masses as an indication of the existence of  rather strong
quark-quark correlations inside baryons, which might help to explain
baryon dynamics. Some lattice simulations reveal  strong attraction
in the diquark channel \cite{wk} with a diquark mass $\sim$600~MeV.
Recently, in \cite{jw}, the mass and extremely narrow width, as
well as other  properties, of the pentaquark  $\Theta^+$  were
explained just on the assumption that it is composed of an antiquark
and two highly correlated $ud$-pairs. At present time, the nature of
the mechanism which may entail strong attraction of quarks in
diquark channels is actively discussed both in the nonperturbative
QCD and in other models (see, \textit{e.~g.\/}\ \cite{ripka} and
references therein).

Finally, let us comment on the fact that we considered in this paper
only a single chemical potential $\mu$ (common for all quarks).
Obviously, in such a simplified approach, the 2SC quark matter is
neither color nor electrically  neutral, as it would be expected for
realistic situations like, for example, in the cores of compact
stars. In order to study color superconductivity in the case of
neutral matter, one has to study more complex NJL models including
several new chemical potentials \cite{neutr}). Despite  this
drawback, the chosen simplified approach, nevertheless, seems to us
interesting enough to get some deeper understanding of the
dynamics of mesons and diquarks in the color symmetric and
2SC phases of cold dense quark matter. A generalization of this
approach to more sophisticated NJL models including several chemical
potentials is presently under investigation.

\section*{Acknowledgments}

We are grateful to D. Blaschke and M. K. Volkov for stimulating
discussions and critical remarks. This work has been supported in
part by DFG-project 436 RUS 113/477/0-2, RFBR grant 02-02-16194, the
Heisenberg--Landau Program 2004, and the ``Dynasty'' Foundation.

\appendix
\section{Loop integrals at $T\ne 0$ and $\mu\ne 0$}

To evaluate loop integrals, we use in our paper the imaginary-time
formalism (see, \textit{e.~g.\/} \cite{kap}). First of all, it is
supposed that the system is in the thermodynamic equilibrium with a
thermal bath of some temperature $T$. As usual, to study a hot and
dense system, one has to calculate thermal Green's functions
(TGF), which are periodic (for each boson field) and antiperiodic
(for each fermion field) functions of imaginary time. Their Fourier
images are, thereby, functions defined at discrete points in the
imaginary axis $p_0=i\omega_n,\, n=0,\pm1,\pm2,\ldots$, with
$\omega_n$ being Matsubara frequencies. When calculating a loop
diagram, one has to sum over $q_0=i\omega_n= i2n\pi T$ for bosons and
$q_0=i\omega_n=i(2n+1)\pi T$ for fermions. In accordance with this,
the integration over $q_0$  in (\ref{140}), (\ref{141}), etc is to be
performed in two steps: First, all momenta are considered in
Euclidean metric, and a sum over Matsubara frequencies is performed.
Then, the limit $T\to 0$ is reached and the resulting TGF are
continued in the complex plane to the real axis, which corresponds to
their definition in Minkowski metric. As a result, we have then for
1PI Green's functions the formulas where the four-dimensional
integration  is reduced to three dimensions.

Let us explain the scheme, briefly outlined above, in a bit more
details. First, we delay the three-dimensional integration until the
end of the calculations, and consider only one-dimensional
$q_0$-integrals. Only one such integration comes from each diagram
in the one-loop approximation for the 1PI Green's functions with two
bosonic legs, and it can be  represented as
\begin{equation}
\label{A1}
F(p_0)= \int\frac{dq_0}{2\pi i}f(q_0,\vec{q};p_0),
\end{equation}
where the function $f(q_0,\vec{q};p_0)$ is a product of vertices and
quark propagators from a one-loop diagram, while the external
three-momentum in each diagram is supposed to be zero, $\vec p=0$.
The dependence of the function $F(p_0)$ on the momentum $\vec{q}$ is
implied, however, we do not write $\vec{q}$ explicitly, since only
the dependence on $p_0$ is important for the moment.

At $T\ne 0$, one would obtain in the right hand side of (\ref{A1})
a sum over the fermionic Matsubara frequencies $\omega_n=(2n+1)\pi T$
instead of the integral. Let us denote the result of the sum
by the function $F_T(p_0)$. It is defined only at the values of
$p_0=i\nu_k=i2k\pi T$, corresponding to bosonic Matsubara
frequencies, because each external line in the implied one-loop
diagram corresponds to a boson. The result  looks like
\begin{equation}
\label{A2}
F_T(i\nu_k)= T\sum_{n=-\infty}^{\infty}f(i\omega_n,\vec{q};i\nu_k).
\end{equation}
Let us assume that  the function $f(\omega,\vec{q};i\nu_k)$ falls
sufficiently quickly in any direction on the complex $\omega$-plane
and contains only simple poles everywhere, except for the imaginary
axis. In this case, using the Cauchy theorem, one can replace the sum
over Matsubara frequencies in (\ref{A2}) by "closed" contour
integrals and obtains
\begin{equation}
  \label{A3}
F_T(i\nu_k)=
T\!\sum_{n=-\infty}^{\infty}f(i\omega_n,\vec{q};i\nu_k)=
  \frac{1}{2}\oint\limits_{C_1+C_2}\frac{d\omega}{2\pi
  i}f(\omega,\vec{q};i\nu_k)\tanh\left (\frac{\omega}{2T}\right ),
\end{equation}
where the contour $C_1$ is just the straight line from $-i\infty
+\epsilon$ to $i\infty +\epsilon$, and $C_2$ is the straight line
from $i\infty -\epsilon$ to $-i\infty -\epsilon$. The contour
integral in (\ref{A3}) is indeed the sum of two integrals: $C_1$ and
$C_2$. Both  $C_1$ and $C_2$  can be closed by  infinite arcs in the
right and left halves of the complex plane because the integration
along these arcs vanishes if the integrands fall quickly enough near
infinity. As a consequence, we can then integrate along the closed
contours $\tilde C_1$ and $\tilde C_2$. As the integrand was supposed
to have only simple poles, the Cauchy theorem immediately gives us
the result of integrations via a sum of residues of the integrand at
these poles, which are determined by the poles of the quark
propagator. It is important to keep in mind before taking the limit
$T\to 0$ that $\nu_k=2k\pi T$. The point is that after the
calculation of all residues the frequency $i\nu_k$ will appear in the
argument of the hyperbolic tangent $\tanh(\omega/2T)$, which is
periodic on the imaginary axis and the period is just equal to $i
2\pi T$. Therefore, the tangent does not depend on $i\nu_k$, and one
can put $\nu_k=0$ in its argument. After this, one can reach the
limit $T\to 0$ and continue the function $F_T|_{T=0}$ to real
energies $p_0$, which is formally obtained through the substitution
$\nu_k\to -i p_0$ in the rest of the expression.

We follow the scheme explained above to calculate  the tadpole
contributions (see (\ref{14})), which do not depend on external
momenta. Let us consider, as an example, the one-loop tadpole
contribution for  $\vev{\Delta^*_2(x)}$, which is nothing else than
the right hand side of (\ref{141}). Replacing the $q_0$-integral by a
sum over Matsubara frequencies $\omega_n=(2n+1)\pi T$ (see
(\ref{15})) and following other steps, we obtain
\begin{eqnarray}
  \label{A5}
I_\Delta(T)&=&-4T\sum_{n=-\infty}^{\infty}\int\frac{d^3q}{(2\pi)^3}
\left\{\frac{1}{D_+(i\omega_n)}+\frac{1}{D_-(i\omega_n)}\right\}=
\nonumber\\
&=&  -2\oint\limits_{\tilde C_1+\tilde C_2}\frac{d\omega}{2\pi
  i}\tanh\left (\frac{\omega}{2T}\right
  )\int\frac{d^3q}{(2\pi)^3}\left\{\frac{1}{D_+(\omega)}+
  \frac{1}{D_-(\omega)}\right\},
\end{eqnarray}
where $D_\pm(\omega)$ are defined by (\ref{13}), and the two
clockwise contours $\tilde C_1$ and $\tilde C_2$
enclose the right and left halves of the complex
$\omega$-plane. The integrand in (\ref{A5}) has only four simple
poles at  $\omega=\pm E^+_\Delta$ and $\omega=\pm E^-_\Delta$.
Finally, we replace the integral in (\ref{A5})
by a sum of the residues at these poles, multiplied by $2\pi i$
(according to the Cauchy theorem), and take the limit $T\to 0$. This
gives for $I_\Delta=\lim_{T\to 0} I_\Delta(T)$ the expression
displayed in the right hand side of (\ref{17}).

\section{Searching for the resonance solution of eq.\ (\ref{32})}

\begin{figure}
\begin{center}
\includegraphics{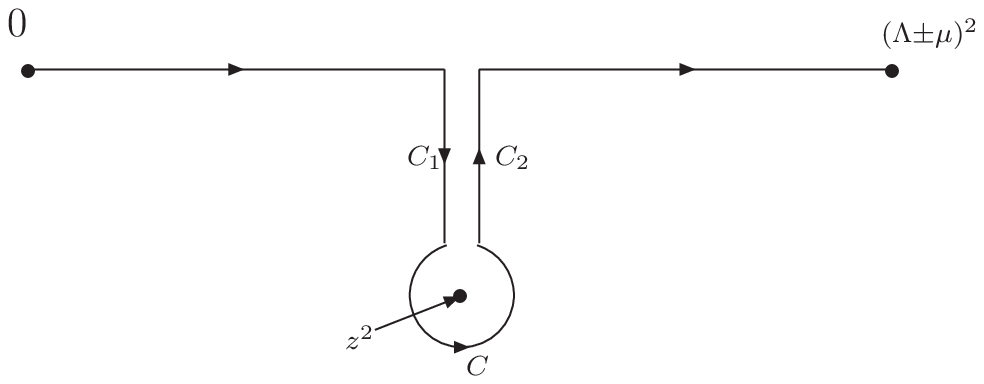}
\end{center}
\caption{The deformation of the integration contour on the complex
$y$-plane for the functions $I_0,I_1$ when the parameter $z^2$
moves onto the lower half-plane of the second Riemann sheet.}
\label{contour1}
\end{figure}

The nontrivial solution of (\ref{32}) obeys the equation
\begin{eqnarray}
(zI_0-2I_1)(zI_0+2I_1)=0,
\label{det1}
\end{eqnarray}
where $z^2=p_0^2-4\Delta^2$ and $I_{0,1}$ are given by (\ref{24}).
Below we shall ignore, for simplicity, the dynamical quark mass
(\textit{i.~e.\/} putting  $m=0$), because we assume (and this
assumption is \textit{a posteriori} corroborated by numerical
calculations of the heavy diquark mass both for $m=0$ and $m\ne 0$)
that this simplification does not strongly affect our results. First
of all, we integrate over the angles in (\ref{25}) and (\ref{26})
and introduce a new variable $y=(E+\mu)^2$, instead of the
three-momentum, in the integrals $A_+$ and $B_+$   (recall that
$E=|\vec q|$)
\begin{eqnarray}
A_+=\frac{1}{4\pi^2}\int_{\mu^2}^{(\Lambda+\mu)^2}\frac
{(\sqrt{y}-\mu)^2dy}
{\sqrt{y}\sqrt{y+\Delta^2}[z^2-4y]},~~~
B_+=\frac{1}{4\pi^2}\int_{\mu^2}^{(\Lambda+\mu)^2}\frac
{(\sqrt{y}-\mu)^2dy}
{\sqrt{y+\Delta^2}[z^2-4y]}.
\label{ABplus}
\end{eqnarray}
Since
\begin{eqnarray}
A_-=\int_0^\mu (\cdots)dE+\int_\mu^\Lambda (\cdots)dE,
\label{Aminus}
\end{eqnarray}
one can introduce the new variables $\sqrt{y}=\mu-E$ and
$\sqrt{y}=E-\mu$ for the first and second integrals of
(\ref{Aminus}), respectively, and get
\begin{eqnarray}
A_-=\frac{1}{4\pi^2}\int^{\mu^2}_0\frac
{(\sqrt{y}-\mu)^2dy}
{\sqrt{y}\sqrt{y+\Delta^2}[z^2-4y]}+\frac{1}{4\pi^2}\int^{(\Lambda-
\mu)^2}_0\frac
{(\sqrt{y}+\mu)^2dy}
{\sqrt{y}\sqrt{y+\Delta^2}[z^2-4y]}
\label{Aminus1}
\end{eqnarray}
and
\begin{eqnarray}
I_0=A_++A_-=\frac{1}{4\pi^2}\int^{(\Lambda+\mu)^2}_0\frac
{(\sqrt{y}-\mu)^2dy}
{\sqrt{y}\sqrt{y+\Delta^2}[z^2-4y]}+\frac{1}{4\pi^2}\int^{(\Lambda-
\mu)^2}_0\frac
{(\sqrt{y}+\mu)^2dy}
{\sqrt{y}\sqrt{y+\Delta^2}[z^2-4y]}.
\label{A}
\end{eqnarray}
In a similar way, one obtains
\begin{eqnarray}
I_1=B_+-B_-=\frac{1}{4\pi^2}\int^{(\Lambda+\mu)^2}_0\frac
{(\sqrt{y}-\mu)^2dy}
{\sqrt{y+\Delta^2}[z^2-4y]}-\frac{1}{4\pi^2}\int^{(\Lambda-
\mu)^2}_0\frac
{(\sqrt{y}+\mu)^2dy}
{\sqrt{y+\Delta^2}[z^2-4y]}.
\label{B}
\end{eqnarray}
The quantities $I_{0,1}$ are analytical functions of the complex
variable $z^2\equiv a-ib$ on the whole complex plane, except for the
cut $L$ on the real axis, defined by $0\leq z^2\leq (\Lambda+\mu)^2$
(the first Riemann sheet). A numerical processing of the integrals
(\ref{A}) and (\ref{B}) gives the values of functions $I_0$ and
$I_1$  on the first Riemann sheet only. But there is no solution for
eq.~(\ref{det1}) in the first Riemann sheet because the root of
eq.~(\ref{det1}) lies on the lower half-plane ($b>0$) of the second
Riemann sheet, and in order to find it, we have to continue $I_0$
and $I_1$ to the second sheet. When $z^2$ crosses the cut $L$
downwards from the upper half of the first sheet, the integrals
$I_0$ (\ref{A}) and $I_1$ (\ref{B}) become singular because of the
apparent poles in the integrands. Since an integral does not change
when the contour is being continuously transformed on the complex
plane until it  crosses a singularity of the integrand, we carefully
deviate our contour on the complex $y$-plane to circumvent the
singularity in a way  shown in Fig.~\ref{contour1}. When the contours
$C_1$ and $C_2$ overlap each other, the sum of the integrals along
them vanishes, and as a result, we obtain that  $I_0$ and $I_1$ in
the lower half-plane of the second Riemann sheet differ from
(\ref{A}) and (\ref{B}) by an additional integral over the contour
$C$ around the pole $z^2$ (see Fig.~\ref{contour1}) which is equal to
the residue of the integrand at $z^2$. Let us denote the
continuations of $I_0$ and $I_1$ to the second Riemann sheet as
$\tilde I_0$ and $\tilde I_1$. Then, one has
\begin{eqnarray}
\tilde I_0=I_0-\frac{i(z^2+4\mu^2)}{4\pi
z\sqrt{z^2+4\Delta^2}},~~~~~~~
\tilde I_1=I_1+\frac{iz\mu}{2\pi\sqrt{z^2+4\Delta^2}},
\label{AB2}
\end{eqnarray}
where  $z=\sqrt{z^2}=a_1-ib_1$, with both $a_1$
and $b_1$ being real and positive: $a_1=\sqrt{\frac 12a+\frac
12\sqrt{a^2+b^2}}$, $b_1=b/\sqrt{2a+2\sqrt{a^2+b^2}}$; apart from,
$\sqrt{z^2+4\Delta^2}=\tilde a_1-i\tilde b_1$, where $\tilde
a_1=\sqrt{\frac 12(a+4\Delta^2)+\frac
12\sqrt{(a+4\Delta^2)^2+b^2}}$,
$\tilde b_1=b/\sqrt{2(a+4\Delta^2)+2\sqrt{(a+4\Delta^2)^2+b^2}}$.

Let us now consider the function
$F(z^2)=\sqrt{z^2}I_0(z^2)+2I_1(z^2)$
(it is the second multiplier in (\ref{det1})). Its continuation to
the second Riemann sheet is $F\to\tilde F(z^2)=\sqrt{z^2}\tilde
I_0+2\tilde I_1=F(z^2)-i(\sqrt{z^2}-2\mu)^2/(4\pi\sqrt{z^2+4
\Delta^2})$. Numerical solution of the equation $\tilde F(z^2)=0$ at
various $\mu>\mu_c$ gives one root per one value of $\mu$:
$z_0^2+4\Delta^2=M^2-iM \Gamma$. For example, if one puts
$\Delta=115$~MeV, $\mu=350$~MeV and $\Lambda=618$~MeV, one gets
the mass $M=1111$~MeV and the width $\Gamma=446$~MeV.

\end{document}